\documentstyle[bbm,amsfonts,epsfig,12pt,here]{article}

\newcommand {\eqref} [1] {(\ref {#1})}
\newcommand {\slsh} [1] {\not{\hbox{\kern-2pt${#1}$}}}

\newcommand {\beq} {\begin{equation}}
\newcommand {\eeq} {\end{equation}}
  \newcommand {\ber}{\begin{eqnarray*}}
  \newcommand {\eer} {\end{eqnarray*}}
\newcommand {\bea}{\begin{eqnarray}}
  \newcommand {\eea} {\end{eqnarray}}

\newcommand{\Dslash}{\,{\raise.15ex\hbox{/}\mkern-12mu D}}
\newcommand{\Tr}{{\rm Tr}\,}
\newcommand{\gsim}{\lower.7ex\hbox{$
\;\stackrel{\textstyle>}{\sim}\;$}}
\newcommand{\lsim}{\lower.7ex\hbox{$
\;\stackrel{\textstyle<}{\sim}\;$}}

\def\beqn{\begin{eqnarray}}
\def\eeqn{\end{eqnarray}}

\begin{document}
\begin{titlepage}
\begin{flushright}{UMN-TH-2624/07\,, FTPI-MINN-07/33\,, SLAC-PUB-13032}
\end{flushright}

\vskip 0.5cm

\centerline{{\Large \bf Planar Limit of Orientifold Field Theories}}

\vskip 0.2cm

\centerline{{\Large \bf and Emergent Center Symmetry }}
\vskip 1cm
\centerline{\large Adi Armoni,${}^{a}$ Mikhail Shifman,${}^{b}$ Mithat \"{U}nsal ${}^{c,d}$}

\vskip 0.3cm
\centerline{${}^a$ \it Department of Physics, Swansea University,}
\centerline{\it Singleton Park, Swansea, SA2 8PP, UK}
\vskip 0.2cm

\centerline{${}^b$   \it William I. Fine Theoretical Physics Institute,}
\centerline{\it University of Minnesota, Minneapolis, MN 55455, USA}

\vskip 0.2cm
\centerline{${}^c$ \it SLAC, Stanford University, Menlo Park, CA 94025, USA}
\vskip 0.1cm
\centerline{${}^d$ \it  Physics Department, Stanford University, Stanford, CA,94305, USA }

\vskip 1cm

\begin{abstract}

We consider orientifold field theories (i.e. SU$(N)$ Yang--Mills theories with
fermions in the two-index symmetric or antisymmetric representations)
on $R_3 \times S_1$ where the compact dimension can be either temporal or spatial. 
These theories are planar equivalent to supersymmetric Yang--Mills.
The latter has $Z_N$ center symmetry.
The famous Polyakov criterion establishing confinement-deconfinement phase transition as that from $Z_N$ symmetric to $Z_N$ broken phase applies.
At the Lagrangian level
the orientifold theories have at most a $Z_2$ center. We discuss how the
full $Z_N$ center  symmetry dynamically emerges in the orientifold theories in the limit
$N\to\infty$.
In the confining phase the manifestation of this enhancement is the existence
of stable $k$-strings  in the large-$N$ limit of the
orientifold theories. These strings are identical to those of supersymmetric 
Yang--Mills theories. We argue  that critical temperatures (and other features)
of the confinement-deconfinement phase transition are the same in
the orientifold  daughters and their supersymmetric parent up to $1/N$ corrections.  
We also discuss the Abelian and non-Abelian confining regimes of four-dimensional 
QCD-like theories.

\end{abstract}

\end{titlepage}

\section{Introduction}
\label{introduction}

In this paper we consider dynamical aspects of four-dimensional orientifold field theories compactified on $R_3 \times S_1$. The compactified dimension  $S_1$ is either 
temporal or spatial. Whether we deal with thermal or spatial formulation of the problem depends on  the  spin connection of fermions along the compact direction. In
the latter case we arrive at a zero temperature field theory where 
phase transitions (if any) are induced quantum-mechanically. 
 In either  case, if the radius of $S_1$
is sufficiently large we return to four-dimensional theory, $R_3 \times S_1\to R_4$.

By orientifold field theories we mean SU$(N)$ Yang--Mills theories with Dirac fermions 
in two-index representations of SU$(N)$ --  symmetric or anti\-symmetric.\footnote{They
will be referred to as orienti-S and orienti-AS, respectively.}
Our starting point is the large-$N$ equivalence between these theories and ${\mathcal N}=1$ 
super-Yang--Mills  (SYM) theory
\cite{Armoni:2003gp,Armoni:2004uu,Armoni:2004ub,Kovtun:2003hr, Kovtun:2004bz} (see also Ref. \cite{Sagnotti} for the relation with string theory). 
Since supersymmetric gauge theories are better understood than nonsupersymmetric, we hope  to learn more about nonsupersymmetric daughters from 
planar equivalence. This expectation comes true:
the center symmetry of SYM theory turns out to be an emergent symmetry of the orientifold daughters in the large-$N$ limit. This fact was first noted in
\cite{Shifman:2007kt} while the first mention of the problem of
center symmetry on both sides of planar equivalence
can be found in \cite{san}. Here we investigate the reasons that lead to
the emergence of  the center symmetry and its implications, 
as they manifest themselves at small and large values of the $S_1$ radius.

For SU(3)  the orienti-AS theory reduces to one-flavor QCD.
If the large-$N$ limit is applicable to $N=3$, at least semi-quantitatively, 
 we can copy SYM theory data
to strongly coupled one-flavor QCD. A concrete example is the temperature independence 
observed in \cite{Kovtun:2007py}. It was shown that certain observables of  SYM
theory are temperature-independent at large $N$ and so is the charge-conjugation-even subset of these observables in the large-$N$ orientifold field theory. It implies a very weak (suppressed by $1/N$) temperature dependence of certain well-defined observables in the confining phase of QCD.
This analytical result is supported by lattice simulations \cite{Lucini:2005vg,Lucini:2004my, Narayanan:2003fc}. For a  recent review, see  \cite{Narayanan:2007fb}.  The planar equivalence 
between SYM theory and oreinti-AS is valid in any phase which does not break the charge conjugation symmetry ($C$ invariance). This implies coinciding Polyakov loop expectation values in the low-temperature confined and high-temperature deconfined phases and the equality of the  the confinement-deconfinement transition temperature
in orienti  and SYM theories in the large $N$ limit.  Other features of the phase transition
are predicted to coincide too.

In the spatial compactification of SYM theory,  the center symmetry is unbroken  at any radius.  For orienti-AS, it is  dynamically broken, along with $C$ and $CPT$, 
at small radii, and restored at a critical radius of the order of $\Lambda^{-1}$ \cite{Unsal:2006pj,Armoni:2007rf}.  
These zero-temperature, quantum phase transitions are observed in recent  lattice simulations by two independent groups \cite{DeGrand:2006qb,Lucini:2007as}. 
The unbroken  center symmetry in the small $S^1$ regime
of the  vector-like gauge theories, unlike the dynamically broken center symmetry, leads to Higgsing of the theory. The long-distance dynamics of such QCD-like theories  are intimately connected to the Polyakov 
model \cite{Polyakov}.  We discuss both the  strong coupling and weak coupling confinement regimes. At weak coupling we get Polyakov (Abelian) confinement which is analytically tractable. The region of validity of the Abelian confinement in 
 QCD-like theories  is a vanishingly small window which diminishes with increasing $N$. 
The fact that the Abelian confinement regime is vanishingly small is a consequence of 
 volume independence.  We discuss this issue in some detail. 
 
\vspace{2mm}

 Summarizing, our findings are:
 
 \vspace{1mm}
 
Orientifold field theories 
on $R_3 \times S_1$ exhibit a number of {\em a priori} unexpected features. 
These theories are planar equivalent to supersymmetric Yang--Mills.
The latter has $Z_N$ center symmetry.
The famous Polyakov criterion establishing confinement-deconfinement phase transition as that from $Z_N$ symmetric to $Z_N$ broken phase applies.
At the Lagrangian level
the orientifold theories have at most a $Z_2$ center. The
full $Z_N$ center  symmetry dynamically emerges in the orientifold theories in the limit
$N\to\infty$.
In the confining phase the manifestation of this enhancement is the existence
of stable $k$-strings  in the large-$N$ limit of the
orientifold theories. These strings are identical to those of supersymmetric 
Yang--Mills theories. The critical temperatures 
of the confinement-deconfinement phase transitions are the same in
the orientifold  daughters and their supersymmetric parent up to $1/N$ corrections.  
Depending on the size of $S_1$ one can identify the Abelian and non-Abelian confining regimes of four-dimensional 
QCD-like theories. 

\vspace{2mm}

The organization of the paper is as follows. In Sect. \ref{planar} we outline
basic facts on
planar equivalence,  define the Polyakov line, and spatial Wilson lines. 
Section \ref{center} is devoted to the center symmetry in SYM theory, QCD with fundamental fermions and orienti-S/AS.  It gives an alternative derivation of  the approximate, exact 
 and emergent center symmetries.  In Sect.~\ref{manifestation} we
discuss 
the strong-coupling manifestation of the emergent center symmetry: existence and
stability of $k$-strings.
In Sect.~\ref{svwcr} we
discuss  strong vs. weak coupling regimes. Section~\ref{p3dc} is devoted
to Polyakov's mechanism of Abelian confinement. In particular, we
address the issue how it  can be generalized to  theories with 
adjoint and two-index  fermions. We confront the thermal 
confinement-deconfinement phase transitions in SYM and orienti theories in Sect.~\ref{thermal}. Here we prove the equality of the critical temperatures at $N\to\infty$. Finally, Sect.~\ref{conclu}
briefly summarizes our results. One-loop potentials are derived in Appendix.
   
\noindent

 \section{ Planar equivalence and Polyakov line}
\label{planar}
\vspace{2mm}

Planar equivalence is equivalence in the large-$N$ limit of
distinct QCD-like theories in their common sectors. 
Most attention received equivalence between SUSY gluodynamics
and its orientifold and $Z_2$ orbifold  daughters.
The Lagrangian of the parent supersymmetric 
theory is
\beq
{\mathcal L}= -\frac{1}{4g_P^2} \, G_{\mu\nu}^a G_{\mu\nu}^a
+\frac{i}{g_P^2}\, \lambda^{a\alpha} D_{\alpha\dot\beta}\bar\lambda^{a\dot\beta}
\label{1}
\eeq
where $\lambda^{a\alpha}$ is the gluino (Weyl) field in the adjoint
representation of SU$(N)$, and $g_P^2$ stands for the coupling constant in the parent theory. The orientifold daughter is obtained by
replacing $\lambda^{a\alpha}$ by the {\em Dirac} spinor in the two-index
(symmetric or antisymmetric) representation (to be referred to as orienti-S
or orienti-AS). The gauge coupling stays intact. 
To obtain the $Z_2$ orbifold daughter  we must pass
to the gauge group SU$(N/2)\times$SU$(N/2)$, replace $\lambda^{a\alpha}$
by a bifundamental Dirac spinor, and rescale 
the gauge coupling, $g_D^2=2g_P^2$ 
\cite{Armoni:1999gc,Strassler:2001fs, Tong:2002vp, Armoni:2005wta, Kovtun:2005kh}. 
We will focus on orienti-AS. Consideration of orienti-S runs in parallel,
with the same conclusions. 

Planar equivalence between the parent and daughter theories  can be applied  in arbitrary geometry, in particular,  on  $S_1\times R_3$, $S_1 \times S_3$ and  $R_4$.
The equivalence implies that correlation functions of the daughter theory are equal 
to the correlation functions of the parent theory at $N\to\infty$. Hence, in 
those cases where the underlying ``microscopic"
symmetries of the planar-equivalent partners do not coincide, the theory with
a  lower microscopic symmetry 
will reflect the symmetries of the parent theory, which are naively absent in the daughter (or vice versa) \cite{Kovtun:2003hr}. The most profound effect of such a 
symmetry mismatch --- and enhancement ---  occurs when  one dimension is compactified onto a circle, and the symmetry under consideration is the center symmetry.  
The corresponding order parameter is the Polyakov line which we define 
below. 
 
Assume that one dimension is compactified (it may be either
time or a spatial dimension). For definiteness, we will assume $z$ to be compactified.
The Polyakov line (sometimes called the Polyakov loop)
 is defined as a path-ordered holonomy of the Wilson line
in the compactified dimension,
\beq
{\cal U} = P\exp\left\{i\int_0^L a_z dz \right\} \equiv V U V^\dagger
\eeq
where $L$ is the size of the compact dimension while
$V$ is a matrix diagonalising ${\cal U}$. Moreover,
\beq
U = {\rm diag}\{ v_1, v_2, ..., v_N\} \equiv e^{ iaL}\,,
\label{3}
\eeq
where
\beq
a =\sum_{\rm Cartan\,\, gen} a_c T^c \equiv {\rm diag}\{ a_1, a_2, ..., a_N\}\,,\qquad  \sum_{k=1}^N a_k =0\,.
\label{4}
\eeq
It is obvious that 
\beq
a_i = -i \ln v_i \,\, {\rm mod}\,\, 2\pi\,.
\eeq
The planar equivalence implies definite relations among the expectation values of the Polyakov loops 
in  SU$(N)$ SYM and orienti theories --- two gauge theories with distinct center symmetries at the Lagrangian level.  In the next section, we will discuss the vacuum structure of these theories  from the center symmetry viewpoint. 

\vspace{2mm}

\vspace{2mm}

\section{Center symmetry (exact and  approximate)}
\label{center}
\vspace{2mm}

In SYM theory all dynamical fields --- gluons and gluinos ---
are in the adjoint representation of SU$(N)$.
This means that the gauge group is
\beq
{\cal G} = {\rm SU}(N)/Z_N
\eeq 
rather than SU$(N)$. This fact manifests itself  as a $Z_N$ symmetry
on the elementary cell
of $\{ a_1, a_2, ..., a_N\}$. Under SU$(N)$ transformations from $Z_N$
\beq
U\longrightarrow e^{\frac{2\pi i k}{N}}\,U\,,\qquad k = 0,1, ..., N-1\,.
\eeq
The $Z_N$ symmetry, usually referred to as 
the  center symmetry, may or may not be spontaneously broken.
There  is a famous Polyakov  criterion regarding
confinement/deconfinement transition in SU$(N)$ Yang--Mills theories.
If one  considers the Polyakov line along the compactified direction, and
its expectation value $\langle {\rm Tr} \, {\cal U}\rangle$ 
does  {\em not} vanish, the center symmetry is broken implying deconfinement.
On the other hand, if $\langle {\rm Tr} \, {\cal U}\rangle = 0$
the center symmetry is unbroken implying 
confinement.\footnote{In QCD-like theories with fermions,  
the boundary conditions on fermions --- antiperiodic vs. periodic --- 
(to be denoted as $\cal S^{\mp}$) determine interpretation of the center symmetry.  If the fermions obey ${\cal S}^{-}$, 
the partition function has a thermal interpretation;
a change in the (temporal) center symmetry realization must be interpreted in terms of 
 the jump in the free energy of the system. If the fermions obey  
 ${\cal S}^{+}$, then the partition function is of the
 twisted type, $\Tr (-1)^F e^{- \beta H}$, and  realization 
 of the spatial center symmetry has interpretation in terms of the  jump in the vacuum energy.}  

Introducing {\em fundamental} dynamical fermions removes  the center symmetry 
(see Ref.~\cite{Weiss:1980rj}). However, one can still make sense 
out of the center symmetry as an approximate symmetry. 
The simplest way to study the impact of the fundamental fermions
is to integrate them out implying the following (formal)  result:
\beq 
\log \det \left(i \gamma_{\mu} D_{\mu}^{\rm F} - m\right) =  \sum_{n \in Z}\,  \sum_{C_n} \alpha(C_n) \, \Tr U(C_n)  
\label{decop}
\eeq
where the superscript F stands for fundamental, 
$\alpha(C_n)$ are 
coefficients scaling with $N$ as $O(N^0)$. The  integer
 $n$ is the winding number of the loop $C$ along the $S_1$ circle which    is valued in the first 
 homotopy group of $S_1$, $$\pi_1(S_1) \sim Z\,.$$ 
A small mass $m$ is inserted 
as an infrared regulator. 
 Note that log of the determinant in  (\ref{decop}),  the fermionic contribution to the action expressed  in terms of the  gluonic observables, scales as $N^1$. 

The $n=0$   sector has  net winding number zero. 
Hence, the corresponding term in (\ref{decop})
is neutral under the center symmetry transformations. 
 However, for instance, the $n=1$ sector
operators are Polyakov loops charged under the center group  transformations.
Thus, the fermion contribution  (\ref{decop}) to the action
is explicitly  non-invariant with respect to the center symmetry. 

A typical term  in the sum from  the winding class $n$   transforms
as 
\beq 
 \Tr U(C_n)  \rightarrow h^n \,  \Tr U(C_n) 
\eeq
where $h \in Z_N$. 
Despite this fact,  the center symmetry is an {\em  approximate symmetry},
since the contribution of the fundamental fermions (\ref{decop}) is suppressed as
$1/N$ relative to the pure glue sector whose action sales as $N^2$.
At $N= \infty$, the fundamental fermions are 
completely quenched and the center symmetry becomes exact. 
The connected correlators or expectation values of the
gluonic observables --- including the Polyakov loop correlators --- 
are the same as in pure Yang--Mills theory.\footnote{The above suppression is analogous to the 
isotopic symmetry in QCD. Since the two lightest flavors are very light compared to the strong scale, 
$m_{u, d}/\Lambda\ll 1$, QCD possesses an approximate SU(2) invariance
despite the fact that the up and down quark masses differ significantly. 
The appropriate parameter $m_{u, d}/\Lambda$ plays the same role as
$1/N$.}

 For orienti-AS  the dynamical AS fermions are {\em not}
 suppressed in the large-$N$ limit.   The above rationale 
 applicable to fundamental fermions no longer holds. However, planar equivalence will lead us to the same conclusion --- emergence of an approximate center symmetry at
 large $N$. 

The Lagrangian of the orientifold theories has the form
\beq
{\mathcal L}= -\frac{1}{4g^2} \, G_{\mu\nu}^a G_{\mu\nu}^a
+\frac{i}{g^2}\, \psi_{ij}^{\alpha} D_{\alpha\dot\beta}\bar\psi^{\dot\beta\,\, ij}
\label{loor}
\eeq
where $\psi_{ij}$ is the Dirac spinor in the two-index
antisymmetric or symmetric representation. Obviously, there is no $Z_N$ symmetry at the Lagrangian level. The center symmetry is $Z_2$ for even $N$ and none for odd $N$.  
 Indeed, the two-index fermion field, unlike that of gluino,
does not stay intact under the action of center elements. The action of a center group element 
on an adjoint fermion is trivial, $\lambda \rightarrow h \lambda h^{\dagger} = \lambda$. The action on an AS fermion is $\psi \rightarrow h\psi h= h^2 \psi$. Thus, for $N$ 
even (odd),   $h=\pm1$ ($h=+ 1$) are the center group elements which leave
the   AS fermion 
invariant, in accordance with at most a $Z_2$ center symmetry  for orienti-AS theory.  

As was mentioned, the antisymmetric fermions are not suppressed 
in the large $N$ limit.  Integrating out the two-index antisymmetric fermion yields 
\begin{eqnarray} 
&& \log \det \left(i \gamma_{\mu} D_{\mu}^{\rm AS} - m\right) 
\nonumber\\[3mm]
&&
=  N^2 \,\sum_{n \in Z} \, \sum_{C_n} \,\frac{\alpha(C_n)}{2}\left\{
\left[\frac{\Tr}{N} U(C_n)\right]^2 -  \frac{1}{N} \frac{\Tr}{N} U^2(C_n)\right\}\,.
\label{Eq:detAS}
\end{eqnarray} 
In the large-$N$ limit we can ignore the single-trace 
terms since they are suppressed by $1/N$
compared to the $O(N^2)$ double-trace term.
The single-trace term contribution scales as that of the
 fundamental fermions, and is quenched in the same fashion. 
   
 A typical double-trace  term $(\Tr U(C_n))^2$ is $O(N^2)$ and is a part of the leading large-$N$ dynamics.  Thus, the impact of the two-index antisymmetric fermions on dynamics is as important as that of  the glue sector of the theory. 
The double-trace term is explicitly  non-invariant under the $Z_N$ center 
transformations. 

We see that in orientifold theories
the center symmetry implementation is much less trivial
than in theories with fundamental quarks. 
As was argued in \cite{Shifman:2007kt},
the center symmetry emerges dynamically in the planar limit $N\to\infty$. 
Here we will carry out a thorough consideration  and  present  independent albeit related
arguments.  

The action of the pure glue sector is local and manifestly invariant under the  $Z_N$ center.  Integrating out fermions, induces a nonlocal sum  (\ref{Eq:detAS}) over gluonic observables.  
This sum includes both topologically trivial loops with no net winding around the 
compact direction  (the $n=0$ term) and nontrivial loops with non-vanishing winding numbers.  
The topologically trivial loops are singlet under the $Z_N$ center symmetry by construction, while  the loops with non-vanishing windings are non-invariant. 

Let us inspect the  $N$ dependence more carefully. 
If we expand the fermion action in the given gluon background  we get
\beq 
\left\langle \exp \left\{- N^2 
\sum_{n \neq 0} \,\sum_{C_n} \frac{\alpha(C_n)}{2}
\left[\frac{\Tr}{N} U(C_n)\right]^2   \right\}    \right\rangle\,,
\label{Eq:exp} 
\eeq
where $\langle ... \rangle$ means averaging
with the exponent combining the gluon Lagrangian with the zero winding number term. 
This weight function is obviously center-symmetric.
If  $h$ is an element of the SU$(N)$ center,  a typical term in the sum (\ref{Eq:exp}) 
transforms as 
\beqn
&& \left\langle \frac{\Tr}{N} U(C_n) \frac{\Tr}{N} U(C_n)\right\rangle  \longrightarrow h^{2n} \left\langle \frac{\Tr}{N}
 U(C_n) \frac{\Tr}{N}  U(C_n) \right\rangle \nonumber\\[4mm]
&&
=
h^{2n}  \left[ \left\langle \frac{\Tr}{N} U(C_n)\right\rangle \left\langle  \frac{\Tr}{N} U(C_n) \right\rangle +   \left\langle \frac{\Tr}{N} U(C_n) \frac{\Tr}{N} U(C_n) \right\rangle_{\rm con}  \right] ,\nonumber\\
 \label{Eq:factor}
 \eeqn 
 where we picked up a quadratic term as an example.
 The connected term in the expression above  is suppressed relative to the leading factorized part 
by  $1/N^2$, as follows from the standard $N$ counting, and can be neglected at large $N$.
As for the factorized part,
 planar equivalence implies that all expectation values 
of multi-winding Polyakov loops 
are suppressed in the large $N$ limit  by $1/N$, 
 \beqn 
&& \left\langle \frac{1}{N} \Tr U (C_n) \right\rangle^{\rm SYM} = 0\,,  
\nonumber\\[3mm]
&&  \left\langle \frac{\Tr}{N} U(C_n) \right\rangle^{\rm AS} = O\left(\frac{1}{N}\right)  \to 0\,,  \quad n \in Z- \{0\}\,,
\label{Eq:lowT}
\eeqn 
where the first relation follows  from unbroken  center symmetry in the SYM theory
and 
the latter is a result of planar equivalence (in the $C$-unbroken,  confining phase of orienti-AS). 

Consequently,  the  non-invariance of the expectation value of the action  under a global center transformation  is  
\beqn 
\langle \delta S \rangle = \langle S(h^n \Tr U(C_n))  -S ( \Tr U(C_n))  \rangle = O\left(\frac{1}{N}\right)  
\langle S \rangle \,,
\label{Eq:noninv}
\eeqn 
which implies, in turn, dynamical emergence  
of center symmetry in orientifold theories in
the large-$N$ limit. 
Let us emphasize again
that the fermion part of the Lagrangian
 which explicitly breaks the $Z_N$ symmetry  is {\em  not} sub-leading in large 
$N$.  However, the effect of the $Z_N$ breaking on  physical observables  is suppressed at
$N\to\infty$. 

This remarkable phenomenon is a natural (and straightforward) consequence of the large-$N$ equivalence  between ${\cal N} =1 $ SYM theory and orienti-AS.  Despite the fact that the center symmetry in the  orienti-AS Lagrangian is at most $Z_2$,  in the $N=\infty$ limit all
observables behave as if they are under the protection of the
$Z_N$ center symmetry. (This point is also emphasized  in \cite{Unsal:2007fb}.) 
We will refer to this  emergent symmetry of the orienti-AS vacuum  as  
 the   {\it custodial symmetry}.    
The custodial symmetry becomes   {\it exact}  in the  $N=\infty$ limit,
 and is {\it approximate} at large $N$.   

 The immediate  implication of this discussion is as follows: when we 
integrate out fermions in the  $N=\infty $  limit, 
the dynamical pattern in orienti-AS/S in the confined phase simplifies. The 
sum over all homotopy classes  in (\ref{Eq:detAS}) reduces to a single term ---
the one over the loops with the vanishing winding number,
$\sum_{n \in Z} \alpha(C_n) \rightarrow  \alpha(C_0)$. Thus, 
\beq 
\log \det \left(i \gamma_{\mu} D_{\mu}^{\rm AS} - m\right)  \sim  N^2
 \sum_{ C_0}
 \frac{\alpha(C_0)}{2} \left[\frac{\Tr}{N} U(C_0)\right]^2 \,.
\label{Eq:detAS2}
\eeq 
Consequently, 
the action and other observables of the $N=\infty$ orienti-AS/S are 
 indistinguishable from the ``reduced" theory with action 
 \beq
 S^{\rm reduced} = S^{\rm YM} +  \sum_{ C_0}
 \frac{\alpha(C_0)}{2} \left[\frac{\Tr}{N} U(C_0)\right]^2 
\eeq
Clearly, the $Z_N$ center is a manifest symmetry of the reduced theory.   Our derivation also 
provides a direct  derivation of the temperature independence  \cite{Kovtun:2007py}  of the orienti-AS theory in the  confining low-temperature phase. 
See also Sect.~\ref{thermal}.

The  $Z_N$-symmetric vacuum structure  at low temperatures, in the $C$
unbroken  phase,
can also be phrased  in terms of the eigenvalue distribution of the Polyakov loop. 
We want to argue that 
the vacuum of the orienti-AS  theory is invariant under the custodial $Z_N$  center symmetry
(which is the symmetry of the  supersymmetric parent).   Let $\rho(\theta)$ denote the distribution of the eigenvalues of the Polyakov loop in orienti-AS  in the 
confined phase. Let us decompose  $\rho(\theta)$ into its Fourier modes, 
$$   
 \rho(\theta) = \frac{1}{2 \pi}\,
\sum_{k \in Z} e^{i \theta k}\, \rho_k  \,.
$$  
All moments  (other than that with $k=0$)  are  restricted  to be  $O(1/N)$   due to 
planar equivalence in the low-temperature phase,
\beqn 
\rho_k =   \left\langle \frac{\Tr}{N} U^k \right\rangle = \frac{1}{N} \sum_{i=1}^{N} e^{i a_i  k} \rightarrow \int d \theta \,
\rho(\theta)\,  e^{i \theta k} \sim \frac{1}{N}\,,\quad k\neq 0\,.
\label{Eq:lowTmode}
\eeqn 
Consequently, in the $N=\infty$ limit,  the eigenvalue distribution of the orientifold theory is flat,
\beq
\rho(\theta)= \frac{1}{2 \pi} \,.
\label{rho2pi}
\eeq
This implies, in particular , that under the action of a center group element, 
$U \to U \,\exp\left\{  \frac{2 \pi i k}{N} \right\} $,  
the eigenvalue distribution 
\beq
\rho(\theta) \to   \rho\left(\theta + \frac{2 \pi  k}{N}\right) =  \rho(\theta) 
\eeq
remains invariant. 

In line with our conclusion are recent calculations on $S_3 \times S_1$ 
\cite{Hollowood:2006cq, Unsal:2007fb} (based on techniques 
developed in \cite{Sundborg:1999ue, Aharony:2003sx})
showing that the vacuum of the large-$N$ 
orientifold theory in the confining phase is characterized by the distribution
(\ref{rho2pi}) i.e. supports the $Z_N$ center 
symmetry rather than the naively expected $Z_2$. 
The authors of  Refs. \cite{ Unsal:2007fb, HoyosBadajoz:2007ds} also reached the conclusion that perturbative transitions which take place on 
$S_3 \times S_1$  capture the nature of the nonperturbative transition taking place in the semi-decompactification  limit of $R_3 \times S_1$, as probed in lattice simulations.

\vspace{2mm}

\section{ Manifestation of $Z_N$ at strong coupling}
\label{manifestation}
\vspace{2mm}

If the circumference of $S_1$ is large enough, $L>L_*$,
 we are in the non-Abelian confinement regime both
in SYM and orientifold theories.\footnote{The value of $L_*$ will be discussed
in Sect.~\ref{svwcr}. Since neither parent nor daughter theories
have massless states in the limit $L\to\infty$, where we recover $R_4$ 
geometry,  the limit must be smooth.} 
 The signature
of the  $Z_N$ center  in SYM theory is the existence of the  $k$-strings.
The tensions and thicknesses of the $k$-strings are class functions of the 
center group $Z_N$. The planar equivalence implies that the orienti-AS theory 
must  have  the very same $k$-strings despite the presence of the dynamical fermions 
charged under the center group. Below we discuss how this arises.

The simplest SYM string is a (chromoelectric)
 flux tube that connects heavy (probe) color sources
in the fundamental representation. Usually it is referred to as the fundamental string.
The  flux tubes attached to color sources
in higher representations of SU($N$) are known as $k$-strings, 
where $k$ denotes the
$n$-ality of the color representation under consideration.
The $n$-ality of the representation with $\ell$ upper and
$m$ lower indices (i.e. $\ell$ fundamental
and $m$ antifundamental)  is defined as 
\beq
k= |\ell -m|\,.
\eeq
It is clear that for stable strings the maximal value of $k=[N/2]$ where 
$[ \; ]$ denotes the integer part. 
The  stability of these $[N/2]$ varieties of strings is a question of energetics. For 
$T_k \leq kT_1$ which is the observation in lattice studies and certain supersymmetric theories, 
all $k$ strings are stable \cite{Lucini:2001nv,Armoni:2006ri}. 

On the other hand, 
in the orientifold-AS theory at finite $N$ (with $N$ even) the only stable $k$-string 
is the fundamental one. If $N$ is odd, there are no stable
$k$-strings at all, as in QCD with fundamental matter \cite{Kratochvila:2003zj}. This is due to the fact that probe charges with even $n$-alities
can be completely screened by two-index antisymmetric quarks,
while those with odd $n$-alities can be screened down up to a single fundamental index
(if $N$ is even) or completely (if $N$ is odd).   
 A similar screening takes
place in the orientifold-S theory with the quark
$\psi_{[ij]}$ replaced by $\psi_{\{ij\}}$.

However, at $N\to\infty$ the breaking amplitude of a color singlet into two is  $1/N$ suppressed. 
Note that the same breaking amplitude in  large-$N$ QCD  with fundamental fermions 
dies off as $1/\sqrt N$. Consequently, 
 in the limit $N\to\infty$
all $k$ strings become stable  against breaking, and identical to those of the SYM theory. The 
$k$-sting  tension, in leading order in $N$, is  $T_k = k T_1$, where  $T_1$ is  the tension of the fundamental string, and is marginally stable. 

Let $W_k (C)$ denote a large Wilson loop in a representation with $n$-ality $k$, with
 $C$ being the boundary  of a surface $\Sigma$.   
The expectation values of such Wilson operators in orienti-AS with odd $N$ is
given by the formula 
\beq
\langle W_k(C)  \rangle_{\rm AS} =  e^{-T_k  A(\Sigma)}  + \frac{1}{N} e^{-\mu_k  P(C)} 
\eeq
where $T_k$ is the  string tension, $ A(\Sigma) $ denotes the area of the surface spanned by the loop  $C$, and $ P(C) $  its perimeter (see Sect.~10 of \cite{Armoni:2003nz}). 
This formula captures two asymptotic
regimes and exhibits noncommutativity of the long-distance and large-$N$ 
limits, a general and quite obvious feature,
\beq
 \lim_{R \rightarrow \infty} \lim_{N\rightarrow \infty} \left[ \frac{\log  \langle W_k(C)  \rangle_{\rm AS}}{R} \right]
   \neq  \lim_{N \rightarrow \infty} \lim_{R\rightarrow \infty} \left[ \frac{\log  \langle W_k(C)  \rangle_{\rm AS}}{R} \right].
\eeq
  At any finite $N$, 
at asymptotically large distances, the perimeter law dominates, and the potential asymptotes to a constant, i.e, 
$$\lim_{R \rightarrow \infty} 
V(R) =  {\rm const}\,.$$
 In this regime, infinite separation of probe color charges  costs
 only a finite energy.
 However, if we take the large-$N$ limit first, linear confinement holds at large distances,  i.e, $$\lim_{R \rightarrow \infty} V(R) =  T_k R\,.$$
Planar equivalence, which implies taking the limit $N \rightarrow \infty$ first, guarantees  equality of the string tensions in  
$N=\infty $ SYM theory  and orienti-AS,
\beq
T_k^{\rm SYM}=  T_k^{\rm AS}\,, \qquad  {\rm for \; all \; }  k,  \qquad N= \infty\,.   
\eeq
We stress again that, unlike fundamental fermions which are quenched in the large $N$ limit, the two-index antisymmetric (symmetric) fermions are not suppressed.  The emergence of $[N/2]$ 
varieties of chromoelectric flux tubes is  associated with the 
suppression of quantum fluctuations in the large-$N$ limit.
This is a nontrivial dynamical effect. 

\vspace{2mm}

\section{ Strong vs. weak coupling}
\label{svwcr}

\vspace{2mm}

SYM theory is planar equivalent to the orientifold theories
provided $C$ (charge conjugation) invariance is not spontaneously broken
\cite{Unsal:2006pj,Armoni:2007rf}.
So far we discussed the large-$L$ limit, i.e. $  L\gg \Lambda^{-1}$ where  $\Lambda$
is the dynamical scale parameter. In this limit both
SYM theory  and its non-supersymmetric daughters are expected to
confine much in the same way as pure Yang--Mills on $R_4$. We will refer to this regime
as {\em non-Abelian} confinement. 

On the other hand, if 
  \beq
  L\ll \Lambda^{-1}\,.
  \eeq
  the gauge coupling is small at the compactification scale. SYM theory with periodic spin connection  (which preserves supersymmetry) undergoes gauge symmetry breaking at the 
  high scale  $\sim   L^{-1}$.  
  The   
  running law of the four-dimensional gauge coupling is changed at the scale where 
   the gauge coupling is still small, i.e. we are in the weak-coupling Higgs regime.
In further descent of the scale the corresponding evolution
of the coupling constant is determined by a three-dimensional theory.

\begin{figure}[h]
 \centerline{\includegraphics[width=2in]{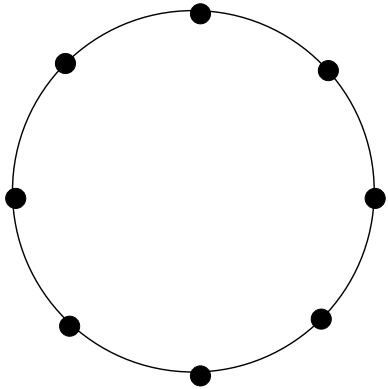}}
 \caption{\small $Z_N$ symmetric vacuum fields $v_k$. For definitions see Eq.~(\ref{3}). }
 \label{zns}
 \end{figure}

In the fully Higgsed regime $a_i\neq a_j $ for all $i, j=1, \ldots N$.  If one chooses 
a generic set of all different $a_i$'s, SYM theory is maximally 
Higgsed; more exactly, SU$(N)$ gauge symmetry is broken down to the maximal
Abelian subgroup U(1)$^{N-1}$. The gauge fields  from the Cartan subalgebra  (as well as the fermions) remain  massless in perturbation theory,\footnote{The nonperturbative mechanism 
 which generates  mass gap $\sim\exp\left\{ - \frac{8\pi^2}{Ng^2}\right\}$ is discussed in 
 Sect.~\ref{p3dc} . }
(they will be referred to as ``photons") while all other gauge fields
acquire masses (they will be referred to ``$W$ bosons").
For generic sets of $a_i$ there is no regular pattern in the $W$ boson masses.
However, if the Higgsed theory is described by $Z_N$-symmetric
expectation values of the diagonal elements $v_k$,
\beq
v_k = e^{\frac{2\pi i k}{N}},\qquad k=1,... , N,
\label{10}
\eeq
(or permutations), see Fig.~\ref{zns},
the  pattern of the $W$ boson masses is regular. $N$ lightest
$W$ bosons (corresponding to simple roots and affine root of SU$(N)$)
are degenerate and  have masses
$\frac{2\pi}{LN}$, while all others can be obtained as
$k\frac{2\pi}{LN}$ where $k$ is an integer.
Thus, there are $\sim N^2$ gauge bosons whose mass 
scales as $1/L$ and $\sim N$ gauge bosons whose mass 
scales as $1/(LN)$.

In the Higgs regime one can consider two distinct sub-regimes.
If \beq
L\lsim L_*\equiv \frac{1}{\Lambda\, N}\,,
\eeq
there exists a clear-cut separation between the scale of the lightest $W$-bosons  
$\frac{2 \pi}{LN}$
 and  nonperturbatively induced photon masses
$\sim\exp\left\{ - \frac{8\pi^2}{Ng^2}\right\}$, where $g$ is the gauge coupling in
the four-dimensional  Lagrangian (\ref{1}) or (\ref{loor}). 
How the nonperturbative photon mass is generated and why it leads
to linear Abelian confinement is explained in Sect.~\ref{p3dc}.
In this regime, the vast majority of $W$ bosons acquire masses that scale as $N $ and, therefore, decouple in the large-$N$
limit. Thus, below the scale $L_*^{-1}$ we deal with three-dimensional
Abelian theory at weak coupling. 
In this theory  Abelian confinement  sets in
due to   a generalization  of the Polyakov mechanism. If the vacuum field is chosen according to (\ref{10})
the Polyakov order parameter vanishes. We are in the $Z_N$-symmetric regime,
much in the same way as in non-Abelian confinement  at $L>\Lambda^{-1}$.
Note that Eq. (\ref{10}) implies that
\beq
\{ a_k L\}  = -\frac{2\pi [N/2]}{N},\,\,  -\frac{2\pi ([N/2]-1)}{N}, ....,\,
\frac{2\pi [N/2]}{N}\,,
\label{12}
\eeq
where $[A]$ stands for the integer part of $A$.

On the other hand, if  we increase $L$, starting from $L_*$ and eventually
approaching $\Lambda^{-1}$, the masses of ``typical"
$W$ bosons become of order $\Lambda$ (the number of such
typical
$W$ bosons $\sim N^2$), while the masses of ``light" $W$ bosons
scale as  $\Lambda /N$ (the number of light
$W$ bosons $\sim N^1$). In this case the effective low-energy description of the theory at $N\to \infty$
must include light $W$'s along with the exponentially
light photons. The expectation value of the  Polyakov loop still vanishes.

 This suggests that the $L$ dependence (in one flavor theories) of   all physical quantities is smooth across the board: from $L\lsim L_*$ to $L\gg \Lambda^{-1}$.
 Rather than a phase
 transition there may be a crossover. 
Theoretical considerations at the moment do not allow
us to prove or disprove the above conjecture. 

The general features of the $L$ behavior are expected to be similar to
those of the $\mu$ evolution of the Seiberg--Witten solution of
${\mathcal N}=2$ Yang--Mills \cite{Seiberg:1994rs}.
(Here $\mu$ is the ${\mathcal N}=2$
breaking parameter in Ref. \cite{Seiberg:1994rs}.)
If $| \mu |$ is small, the
Seiberg--Witten solution applies exhibiting Abelian confinement.
As  $| \mu |$ evolves to larger values, eventually becoming
$\gg \Lambda$, non-Abelian confinement is expected to set in.
Since $\mu$ is a holomorphic parameter one does not expect to have
a phase transition on the way; the $\mu$ evolution
is expected to be smooth.

The
nature of the transition between Abelian and  non-Abelian 
confinement  a good question for lattice studies. In terms of the expectation value of the Polyakov loop,  
$\langle \Tr U \rangle=0$ in all regimes. However, at small $L$ , this is due to the fact that the
eigenvalues of the Polyakov line are, at zeroth order, frozen at the roots of unity 
as in Fig.~\ref{zns} (Higgsing), and  fluctuations are negligibly small. 
At large $L$,  these eigenvalues are strongly coupled and  randomized over the
$[0, 2 \pi)$ interval. Consequently, there is no gauge symmetry breaking
(non-Abelian confinement).  

The set of fields in Eq. (\ref{12}) is automatically invariant under the
$C$ trans\-formation. Therefore, planar equivalence
between SYM  and orientifold theories must hold both
in the Higgs regime, and at strong coupling
where $C$ invariance was argued to hold too \cite{Armoni:2007rf,Shifman:2007kt}.

We would like to stress that the set of fields  in Eq. (\ref{12}) 
should be considered, for the time being, as a fixed background field configuration.
Generally speaking, it does not minimize the energy functional. For instance, in
the thermal compactification this field configuration realizes the
maximum of the effective potential, rather than the minimum.
To get the set of fields Eq. (\ref{12}) as a vacuum configuration
(i.e. that minimizing the effective potential)
we have to change the pattern of compactification (e.g. $S_3 \times S_1$)
or introduce a deformation of the theory through
addition of source terms or both.  What is important for us here is that
this is doable and it is perfectly
reasonable to quantize the theory in the 
$Z_N$ invariant background (\ref{12}) which realizes maximal Higgsing.

Then, it is instructive to illustrate how planar equivalence between
SYM and orientifold theories works in this regime 
by  examining the one-loop example. Given the background fields (\ref{4})
the effective potential for the SYM theory is 
\begin{eqnarray}
   V_{\rm eff} &=& \frac{1}{24\pi^2} \>
    \Big\{
      \sum_{i,j=1}^{N}
	  [a_i {-} a_j ]^2 \left( 2\pi - [a_i {-} a_j] \right)^2
	- \frac{8}{15} \, \pi^4 N
\nonumber\\ && \qquad{}
      - 2 \sum_{i<j=1}^{N}
	  [a_i {-} a_j ]^2 \left( 2\pi - [a_i {-} a_j] \right)^2
    \Big\} \,,
\label{sympotent}
\end{eqnarray}
where everything is measured in the units of $L$.
The effective potential for the orientifold-AS theory (see Appendix)
\begin{eqnarray}
   V_{\rm eff} &=& \frac{1}{24\pi^2} \>
    \Big\{
      \sum_{i,j=1}^{N}
	  [a_i {-} a_j ]^2 \left( 2\pi - [a_i {-} a_j] \right)^2
	- \frac{8}{15} \, \pi^4 N
\nonumber\\ && \qquad{}
      - 2 \sum_{i<j=1}^{N}
	  [a_i {+} a_j ]^2 \left( 2\pi - [a_i {+} a_j] \right)^2
    \Big\} \,,
\label{oripotent}
\end{eqnarray}
For the $Z_N$ symmetric background (\ref{4})
the expressions (\ref{sympotent}) and (\ref{oripotent})
are identical up to terms suppressed by powers of $1/N$.

\section{ Polyakov's 3D confinement}
\label{p3dc}

\vspace{2mm}

Long ago Polyakov considered three-dimensional SU(2) Georgi--Glashow mo\-del 
(a Yang-Mills adjoint Higgs system)  
in the Higgs regime \cite{Polyakov}. In this regime SU(2) is broken down to U(1),
so that at low energies the theory reduces to compact electrodynamics. 
The dual photon is a scalar field $\sigma$ of the phase type
(i.e. it is defined on the interval $[0, 2\pi ]$):
\beq
F_{\mu\nu}
 =\frac{g_3^2}{4\pi} \, \varepsilon_{\mu\nu\rho}\left( \partial^\rho\,\sigma\right)\,,
 \label{14.8}
\eeq
where  $g_3^2$  is the three-dimensional  gauge coupling with mass dimension
  $ [g_3^2]=+1$. 
 In perturbation theory the dual photon $\sigma$
is massless.   However, it acquires a mass
due to instantons (technically, the latter are identical to the 't Hooft--Polyakov
monopoles, after the substitution of one spatial dimension by imaginary time).
In the vacuum of the theory,  one deals with a gas  of instantons interacting
according to the Coulomb  law.   
The dual photon mass is due to the Debye screening.
In fact, it is determined by the one-instanton vertex,
\beq
m_\sigma \sim m_W^{5/2} g_3^{-3}e^{-S_0/2}
\eeq
where $S_0$ is the one-instanton action,
\beq
S_{0} = 4\pi \, \frac{m_W}{g^2_3}\,,
\label{14.5}
\eeq
$m_W$ is the $W$ boson mass. 
As a result, the low-energy theory
is described by a three-dimensional sine-Gordon model,
\beq
{\cal L}_\sigma = \frac{g_3^2}{32\pi^2} (\partial_\mu\sigma )^2 + c_1 
m_W^5g_3^{-4}e^{-S_0}
 \, \cos\sigma\,. 
\eeq
where $c_1$ is an undetermined prefactor. 
This model supports a domain line\,\footnote{Similar to the axion domain wall.} (with 
$\sigma$ field vortices at the endpoints)
which in 1+2 dimensions must be interpreted as a string.
Since the $\sigma$ field  dualizes three-dimensional photon, the $\sigma$ field vortices
in fact represent electric probe charges in the original formulation, connected by the
electric flux tubes which look like domain lines in the dual formulation. 

As well-known \cite{Affleck:1982as}, addition of  one (or more)   
Dirac fermions in the adjoint representation  
eliminates the above Abelian confinement in  the Polyakov model. This is due to the fact
that the instanton-monopole acquires fermion zero modes  due to  the Callias index theorem \cite{Callias:1977kg, Jackiw:1975fn}. 
Instanton-monopoles,  instead of generating mass for the the $\sigma$ field 
through the potential $e^{i \sigma}  + e^{-i \sigma}$,    produce a  vertex with compulsory fermion zero modes attached to it. For 
one flavor theory, this is given by 
\beq
e^{-S_0}(e^{i \sigma} \psi \psi + e^{- i \sigma} \bar \psi \bar \psi          )\,.
\label{manif}
\eeq
 The three-dimensional microscopic theory  with an adjoint fermion possess a non-anomalous 
 U(1)  fermion number symmetry.  
 This  U(1)  invariance  
is manifest in Eq.~(\ref{manif}); it intertwines the  fermion global rotation with a continuous shift symmetry for the dual photon,
\beq 
\psi \rightarrow  e^{i \alpha} \psi, \; \; \;  \sigma \rightarrow \sigma- 2 \alpha \,.
\eeq
 The  continuous shift symmetry  (unlike a discrete one)  prohibits any mass term 
 (or potential) for the $\sigma$ field. As was shown in \cite{Affleck:1982as},
the  U(1)  fermion number is spontaneously broken, and 
the dual photon $ \sigma$  is the associated  Goldstone particle. 
Thus, the $\sigma$ field remains massless nonperturbatively,
and linear confinement does not occur. This is because domain lines become infinitely thick  
(infinite range) in
the absence of the dual photon mass. One does need a 
non-vanishing dual photon mass
to make the domain line thickness of the order of $m_\sigma^{-1}$.
Only then, at distances $\gg m_\sigma^{-1}$ linear confinement will set in.

As was noted in  \cite{Unsal:2007vu},
this obstacle is circumvented if we consider a three-dimensional model obtained as a low-energy reduction of the four-dimensional model compactified on $S_1\times R_3$.
The adjoint Weyl fermion in four dimensions
becomes an adjoint  Dirac fermion in three  dimensions.  In this case, there is no 
U(1) fermion number symmetry. There is an anomalous U(1)$_A$; because of the 
anomaly only a discrete subgroup of U(1)$_A$ is a valid symmetry.  For SYM, the anomaly free 
subgroup is $Z_{2N, A}$.
As stated earlier, the discrete shift symmetry does not
prohibit a mass term for the dual photon, hence it must be  generated \cite{Unsal:2007vu}.  
Whatever non-perturbative object is responsible for
 the dual photon mass, it should have 
no fermionic zero mode ``attached." Otherwise, it will generate vertices as in  (\ref{manif}) 
which do not result in the bosonic potential. 

The microscopic origin of the mass term can be traced to the compactness of the adjoint Higgs 
field  whenever  we consider a theory on $S_1 \times R_3$.  This is the feature which is absent in the Polyakov model \cite{Polyakov} and its naive  fermionic extension \cite{Affleck:1982as}. 
When the adjoint Higgs field is compact as in Fig.\ref{zns}, 
in  additional to $N-1$ 't Hooft-Polyakov monopole-instantons (whose existence
is tied up to $\pi_1 (S_1) \neq 0$) there is one extra, which can be referred to as
the Kaluza--Klein (KK) monopole-instanton.\footnote{The eigenvalues 
shown in Fig.~\ref{zns} may be viewed as Euclidean D2-branes.   $N$ split branes support a spontaneously broken 
U(1)$^{N}$ gauge theory, whose  U(1)  center of mass decouples, and the resulting theory is 
U(1)$^{N-1}$.   The $N-1$  BPS monopoles may be viewed as the Euclidean D0 branes connecting eigenvalues $(a_1 \rightarrow  a_2), (a_2 \rightarrow  a_3), \ldots, (a_{N-1} \rightarrow 
a_{N})$. Clearly, we can also 
have a monopole which connects $(a_N \rightarrow  a_1)$ which owes its existence to the periodicity of the adjoint Higgs field,  or equivalently,  to the fact that the underlying theory is on $S_1 \times R_3$.  Usually it is  called the KK monopole.  The 
Euclidean D0 branes with the
opposite orientation, connecting $(a_{j} \leftarrow a_{j+1}),\,\, j=1, \ldots 
N $ are the antimonopoles.  This viewpoint makes manifest 
the fact that the KK  and 't Hooft--Polyakov monopoles are all on the same footing.  The magnetic and topological charges of the monopoles 
connecting  $(a_{j} \leftrightarrow a_{j+1}) $ is 
$\pm \Big( \mbox{ \boldmath $\alpha$}_j, \frac{1}{N} \Big)$ 
where the direction of the arrow is correlated with the sign of the charges. 
}
Each of these monopoles carries two zero modes, hence they cannot contribute to the bosonic  potential. 
The bound state of the 't Hooft--Polyakov monopole-instanton with magnetic charge  
 $\mbox{\boldmath $\alpha$}_i$ and anti-monopole with charge  $-\mbox{\boldmath $\alpha$}_{i+1}$  has no fermion zero mode: in the sense of topological charge, it is indistinguishable 
 from the perturbative vacuum.
  Hence, such a bound state can contribute to the bosonic potential.
If we normalize the magnetic and topological  charges of the monopoles  as 
\beq
\left(\int  F, \,\,\int F \tilde F \right) = \left( \mbox{\boldmath $\alpha$}_i ,\,\, \frac{1}{N} \right) ,
\eeq
where $ \mbox{\boldmath $\alpha$}_i$ stand for roots of the affine Lie algebra,   then the
following bound states are relevant:
\beq
\left( \mbox{\boldmath $\alpha$}_i,\,\, \frac{1}{N} \right)  + \left( - \mbox{\boldmath $\alpha$}_{i+1}, \,\,- \frac{1}{N} \right) =\left( \mbox{\boldmath $\alpha$}_i - \mbox{\boldmath $\alpha$}_{i+1}, \,\, 0 \right).
\label{38}
\eeq
 This pair is  stable,  as was shown in Ref.~\cite{Unsal:2007vu}, where it is referred as 
a  magnetic bion.    
 Thus,  we can borrow Polyakov's discussion of magnetic monopoles and apply directly 
 to these objects.   The magnetic bions  will induce a mass term  for the dual photons 
 via the Debye screening, the essence  Polyakov's mechanism. 
 
 In the SU$(N)$  gauge theory on $R_3\times S_1$, which
 is Higgsed, ${\rm SU}(N) \rightarrow U(1)^{N-1}$, the bosonic part of the effective low-energy Lagrangian is generated by the pairs 
 (\ref{38}), and hence the potential is proportional to $e^{-2S_0}$,
 rather than $e^{-S_0}$ in the Polyakov problem. 
If we introduce an $(N-1)$-component  vector $\mbox{\boldmath $\sigma$}$,
\beq 
\mbox{\boldmath $\sigma$} \equiv \left(\sigma_1, ...., \sigma_{N-1}\right), 
\eeq 
representing $N-1$ dual photons
of the $U(1)^{N-1}$ theory, and $\mbox{\boldmath $\alpha$}_i$ ($ i=1, ... , N$) are the simple and affine roots of the 
SU$(N)$  Lie algebra, the bosonic part of the effective Lagrangian can be written as
\beq
{\cal L}(\sigma_1, ...., \sigma_{N-1}) =  \frac{g_3^2}{32\pi^2} (\partial_\mu\mbox{\boldmath $\sigma$} )^2 +
 c m_W^5g_3^{-4}e^{-2S_0} 
 \, \sum_{i=1}^{N} \cos \left( \mbox{\boldmath $\alpha$}_i - \mbox{\boldmath $\alpha$}_{i+1}\right)\mbox{\boldmath $\sigma$}\,,
 \label{40}
\eeq
where $c$ is an undetermined coefficient
and $g_3$ is the three-dimensional coupling constant,
\beq
g_3^2 = g^2\, L^{-1}\,.
\eeq
We remind  that $\mbox{\boldmath $\alpha$}_i$ ($ i=1, ... , N-1$) represent the magnetic charges of $(N-1)$ types of the
't Hooft-Polyakov monopoles while the affine root 
\beq
\mbox{\boldmath $\alpha$}_N= -\sum_{i=1}^{N-1} \mbox{\boldmath $\alpha$}_i
\eeq
 is the magnetic charge 
of the KK monopole.  
 Note that the configurations  that contribute to the effective Lagrangian 
have magnetic charges $\mbox{\boldmath $\alpha$}_i - \mbox{\boldmath $\alpha$}_{i+1}$ and vertices  $e^{i(\mbox{\boldmath $\alpha$}_i - \mbox{\boldmath $\alpha$}_{i+1}) \mbox{\boldmath $\sigma$}}$,  corresponding to a  product of a  monopole vertex  
 $e^{i\mbox{\boldmath $\alpha$}_i \mbox{\boldmath $\sigma$}}$
 with charge  $\mbox{\boldmath $\alpha$}_i$,  and   antimonopole 
 vertex $e^{-i\mbox{\boldmath $\alpha$}_{i+1} \mbox{\boldmath $\sigma$}}$
  with charge $-\mbox{\boldmath $\alpha$}_{i+1}$.  We used Eq. (\ref{12}) to guarantee that
the vacuum configuration is $Z_N$-symmetric, hence  the actions (fugacities)  $e^{-2S_0}$ 
are all equal.   

Equation (\ref{40})
implies that non-vanishing 
 masses are generated for all $\sigma$,   proportional to
$e^{-S_0}$,  albeit  much smaller than the masses in 
the Polyakov model in which they are    $\sim e^{-S_0/2}$. 
  There are $N-1$ distinct U(1)'s in this model, corresponding
to $N-1$ distinct electric charges. These are the electric charges $q_i$
of all color components of
a probe non-dynamical quark $Q_i$ ($i=1, ... , N$) in the fundamental representation of 
SU$(N)$.\footnote{$q_i$ are subject to the condition
$\sum_{i=1}^N q_i =0$.}
Correspondingly, there are
$N-1$ types of Abelian strings (domain lines). Their tensions are equal to each
other and proportional to $e^{-S_0}$. Linear confinement develops
at distances larger than $e^{S_0}$.

Needless to say,  the physical spectrum
in the Higgs/Abelian confinement regime is much richer than that
in the non-Abelian confinement regime. If in the latter case
only color-singlets act as asymptotic states, in the Abelian confinement regime
all systems that have vanishing $N-1$ electric charges have finite mass and represent
asymptotic states.

\vspace{2mm}

\section{Thermal compactification}
\label{thermal}
\vspace{2mm}

As was already mentioned the requirement for planar equivalence  to hold is to have a 
vacuum with an unbroken $C$ conjugation symmetry. 
In the non-Abelian confinement regime $C$ parity is unbroken.
In the Higgs regime this depends on the choice of $v_k$'s. 
The set (\ref{10}) automatically guarantees $C$ parity.
Now we abandon this choice and will focus on the case
$a_k=0,\,\,\pi$ relevant to the thermal compactification. The vacuum field 
$a_k=0$ or $a_k=\pi$ for all $k$ is not $Z_N$ symmetric while,
as we already know,  the non-Abelian confinement regime
is $Z_N$-symmetric. Hence, we expect a phase transition 
at a deconfinement temperature $T_*$. 

When the temperature is high,
namely $T>T_*$ (the deconfining phase), the Polyakov loop expectation value
in orienti-AS/S
does not vanish; the minimal energy states are $C$ preserving vacua $a_k=0$ or $a_k=\pi$ for all $k$.
If we choose the same minima of the effective potential  in the high-temperature phase
of the SYM theory,  planar equivalence will hold.  Namely, the SYM theory and orienti-AS/S
will  have the same Green functions, condensates, spectra, etc. in the common sector.

Since all the eigenvalues of the Polyakov line coincide in the high temperature phase, say at
$a_k=0$ for all $k$,  the high-temperature theory is not Higgsed.
The would-be $W$ bosons remain massless.
Its dynamics is that of the non-Abelian theory, albeit three-dimensional rather than
four-dimensional. It  is  believed that the phase transition at $T_*$ separates
the $Z_N$ symmetric phase from that with broken $Z_N$
in pure Yang--Mills or SYM theory.
The orientifold theories seemed puzzling from this standpoint \cite{san}.
Now we understand that the phase transition at $T_*$ in orienti-AS/S
is quite similar: it separates the high-temperature phase with
no $Z_N$ center symmetry (at best, it is $Z_2$ for even $N$ which is spontaneously broken) from the low-temperature phase with emergent unbroken $Z_N$.

 The physical 
 observables in the common sector of SYM theory and orienti-AS/S must coincide throughout the  whole phase diagram, since charge conjugation symmetry is unbroken in neither 
 phase of the theory.  This implies, 
 in particular, the phase transition temperatures must coincide,
 \beq
 T_*^{\rm SYM} =  T_*^{\rm orienti\mbox{-}AS/S}\,, \qquad N\to\infty\,.
 \eeq
The same applies to the  Polyakov order parameter.   At $N=\infty$,   
\beqn 
\left\langle \frac{1}{N} \, \Tr U \right\rangle^{\rm SYM,\,orienti\mbox{-}AS/S } = \left\{ \begin{array}{ll} 
                                             0, \;  & T < T_* \\
                                             \pm 1 , \;   & T >T_*
                                             \end{array} \right.
                                            \label{Eq:deSYM}
\eeqn 
where the sign double-valuedness
corresponds to two possible vacuum fields, $a_k=0, \,\,\forall \, k$ or $a_k=\pi, \,\,\forall \, k$.
(In SYM theory these two minima correspond to the ones invariant under the naive $C$, where the 
comparison is straightforward.) In other words, the notions of high and low temperatures, defined relatively  the deconfinement transition temperature must coincide 
in the two theories. Otherwise we would get an inconsistency among the common sector observables.  This gives us a nontrivial dynamical result,  at finite $N$,
\beq
T_*^{\rm SYM}= T_*^{\rm orienti\mbox{-}AS/S}\left[1 + O\left(\frac{1}{N}\right)\right]  .
\eeq
 This prediction of planar equivalence should be easily  testable on   lattices. 
 
 \section{Conclusions}
 \label{conclu}
 
 The strong coupling dynamics of  non-supersymmetric vector-like gauge theories, 
 despite many efforts over the years, remains elusive.  Currently, the nonperturbative large-$N$ 
 equivalences 
 provide  deep hints   into the  structure of the  vector-like gauge theories.  
 One of the most profound examples of the large-$N$ equivalences is that
 between ${\mathcal N}=1$ SYM theory and its orientifold daughters.
The parent and daughter theories   are clearly distinct, with different fundamental symmetries 
 and dynamics.  At the Lagrangian level, 
  SYM theory is  supersymmetric and has $Z_N$ center symmetry, 
while orienti theories are non-supersymmetric and have at most a $Z_2$  center. 

  However,  the large-$N$ equivalence tells us that  these theories become indistinguishable in their neutral sector in the large-$N$ limit.   More specifically, it tells us that the physical Hilbert spaces of these two theories in the   $C$-even subsectors coincide,  their  confinement-deconfinement temperatures are identical, and the $k$-string tensions must match. 
  
  Apparently, 
 all these observables are associated with certain symmetries which are explicit in 
 the parent  theory, but not in the daughter ones. In the large-$N$ limit the  correlators of the daughter theories carry benchmarks of the custodial symmetries of its parent.   We refer  to such symmetries, 
 which are absent at the Lagrangian level but appear dynamically in the neutral
 correlators, as to {\em emergent symmetries}. In orienti-AS, the $Z_N$
 center symmetry and supersymmetry 
(protecting the degeneracy of the bosonic spectrum)  are emergent symmetries in  this sense. 

At the level of the Schwinger--Dyson equations (or loop equations),  
the equivalence is a consequence 
of the quantum fluctuation suppression  in the large-$N$ limit.  The (suppressed) fluctuations are well aware of distinctions in the
parent/daughter theories, which have no place in the leading large-$N$ dynamics.   This is  true for  orienti-AS/S,  as well as for orbifold SU$(N)\times$SU$(N)$
Yang--Mills with bifundamental quark
 (assuming unbroken $Z_2$ 
interchange symmetry). This is also valid  for one-flavor QCD  with orthogonal and symplectic gauge groups 
with AS/S representation fermions \cite{Armoni:2007jt}.  
 In our opinion, attempts to understand such  universal behavior  (which is a natural consequence of planar equivalence) may provide insights  into the strongly coupled regimes of QCD, and other strongly coupled systems.   This question is left for the future work.

\section*{Appendix: One-loop potentials}
\renewcommand{\theequation}{A.\arabic{equation}}
\setcounter{equation}{0}

The one-loop effective potentials for Polyakov lines in the case of SYM theory
and orienti-AS are
\beqn
&&V_{\rm eff}^{\rm SYM} [U]=  \frac{2}{\pi^2 L^4} \sum_{n=1}^{\infty} \frac{1}{n^4} \left[ (- 1 + a_n)
 |\Tr U^n|^2 \right] \nonumber\\[3mm]
&&V_{\rm  eff}^{\rm A} [U]=  \frac{2}{\pi^2 L^4} \sum_{n=1}^{\infty} \frac{1}{n^4} \left\{ - |\Tr U^n|^2  +  {a_n}  \left[ \frac{(\Tr U^n)^2 -  \Tr U^{2n}}{2} + {\rm c.c.} \right]   \right\}
\nonumber\\
\eeqn
where the first terms are due to gauge boson (and ghosts), and second term is due to fermions endowed 
with spin structure  
\beq 
a_n= \left\{ \begin{array}{cl}
                      (-1)^n  & {\rm for}\;  {\cal S}^{-}\cr
                       1  & {\rm for} \; {\cal S}^{+}\
                       \end{array}
                       \right.                      
\eeq 
In the large $N$ limit,    the single trace term can be neglected since it is suppressed by  $O(1/N)$ relative  to the double trace terms.  In terms of the simultaneous eigenstates of the $C$ and  center symmetry, 
\beq
\Tr \Omega_{\pm}^{n} = \Tr U^{n} \pm \Tr (U^{ *})^n, \qquad  C \Tr \Omega_{\pm} = \pm \Tr \Omega_{\pm},  
\eeq 
 the potential takes the form 
\beqn
 &&V_{\rm  eff}^{\rm AS} [\Omega_+, \Omega_{-}]=  \frac{2}{\pi^2 L^4} \sum_{n=1}^{\infty} \frac{1}{4 n^4} \left[ (-1+ a_n)
  |\Tr \Omega_{+}^n|^2  +    (-1- a_n) |\Tr \Omega_{-}^n|^2            \right] .
  \nonumber\\  
\eeqn
The form of the one-loop potential in QCD(AS) is 
not surprising.  The first class  of 
terms $|\Tr \Omega_+^{n}|^2$  are  the images 
of the $|\Tr U|^2$.   The second category are the square of the twisted (non-neutral) $C$-odd operators, 
which are not the image of any operator in the orienti-partner.  They do, however,  get induced 
via a one loop Coleman--Weinberg analysis.    Whether  or not the twisted operator  
$\Tr \Omega_{-}$ acquires a vacuum expectation value and  induces the spontaneous 
braking of charge conjugation is correlated with the spin structure ${\cal S}^{\pm} $ 
of fermions along the $S_1$ circle.   In thermal case   (${\cal S}^{+} $),  despite the presence of 
such operators,  $C$ is unbroken at high temperature. In spatial compactification  (${\cal S}^{-} $),   $C$ is broken at small $S_1$.  Regardless of spin structure, $C$ is preserved at large radius (either temporal or spatial)  continuously connected to $R_4$.  
 
 \vspace {2mm}

\section*{Acknowledgments}

We are grateful to Larry Yaffe for discussions. A.A.~is supported by the PPARC advanced fellowship award.
The work of M.S. is supported in part by DOE grant DE-FG02-94ER408. 
The work of M.\"U.  is  supported by the
U.S.\ Department of Energy Grants DE-AC02-76SF00515.

\end{document}